\documentclass[10pt]{iopart}
\usepackage{graphicx}
\usepackage{subfigure}
\usepackage{morefloats}
\usepackage{color}
\usepackage{setspace}
\usepackage{floatrow}
\usepackage{amsmath,bm}
\begin{document}

\title[Dynamics of thermal filaments]{\textbf{Dynamics of 3D isolated thermal filaments}}
\author{ N. R. Walkden$^{1}$, L. Easy$^{1,2}$, F. Militello$^{1}$ and J. T. Omotani$^{3}$
        \\ \small{$^{1}$ EURATOM/CCFE Fusion Association, Culham Science Centre, Abingdon, OX14 3DB, UK} 
        \\ \small{$^{2}$ York Plasma Institute, Department of Physics, University of York, Heslington, York, YO10 5DD, UK} 
        \\ \small{$^{3}$ Department of Physics, Chalmers University of Technology, 412 96 Gothenburg, Sweden}
        \\ Email: \texttt{nick.walkden@ccfe.ac.uk} }
\date{}

\begin{abstract}
Simulations have been carried out to establish how electron thermal physics, introduced in the form of a dynamic electron temperature, affects isolated filament motion and dynamics in 3D. It is found that thermal effects impact filament motion in two major ways when the filament has a significant temperature perturbation compared to its density perturbation: They lead to a strong increase in filament propagation in the bi-normal direction and a significant decrease in net radial propagation. Both effects arise from the temperature dependence of the sheath current which leads to a non-uniform floating potential, with the latter effect supplemented by faster pressure loss. The reduction in radial velocity can only occur when the filament cross-section loses angular symmetry. The behaviour is observed across different filament sizes and suggests that filaments with much larger temperature perturbations than density perturbations are more strongly confined to the near SOL region. 
\end{abstract}

\section{Introduction}
Filaments are coherent, meso-scale plasma structures that are strongly aligned to the magnetic field\cite{ZwebenPoP2002,BoedoPoP2001,BoedoPoP2003,BoedoReview,MaquedaRSI2001,KirkPPCF2006,DudsonPPCF2008,AyedPPCF2009,D'IppolitoReview}. They are  universally observed in the boundary plasma of tokamaks \cite{D'IppolitoReview} and other magnetically confined plasmas \cite{AntarPoP2003,KatzPRL2008,TheilerPoP2011} and can play an important role in cross-field transport into and within the scrape-off layer \cite{BoedoPoP2003,MilitelloPPCF2013}. Since filaments are non-local intermittent events their transport cannot be captured by local transport theory \cite{GarciaJNM2007,NaulinJNM2007} so modelling of filamentary transport tends to rely on numerical simulation in 2D \cite{YuPoP2003,YuPoP2006,OmotaniPPCF2015} and, more recently in 3D in both a simple slab \cite{AngusPRL2012,AngusPoP2012,AngusPoP2014,EasyPoP2014,EasyPoP2016} and in more complex geometries \cite{WalkdenPPCF2013,HalpernPoP2014,WalkdenNF2015}. Since the first description of filament motion \cite{KrashenninikovPLA2001} the vast majority of filament dynamics modelling has been carried out assuming an isothermal plasma in order to reduce the complexity of the system under study. It is unlikely however that turbulent processes that produce filaments are isothermal in nature with 2D edge turbulence simulations showing fluctuations in both density and temperature \cite{MilitelloPPCF2013}. This implies that filaments may carry significant temperature perturbations as well as density perturbations into the SOL. It is therefore important to establish what effects the additional temperature perturbation can have on the motion of the filament beyond isothermal dynamics. Such effects have been captured heuristically in reduced 2D models of isolated filaments \cite{D'IppolitoPoP2002,MyraPoP2004} where the electron temperature was shown to produce a monopolar electrostatic potential that can spin the filament on its axis. This occurs as the hotter portion of the filament induces an increase in the floating potential which then biases the filament with a monopolar profile, leading to circulatory $\textbf{E}\times\textbf{B}$ motion.  However 2D models inherently approximate parallel dynamics with heuristic closures in the parallel direction \cite{EasyPoP2014} so a full 3D treatment is required for parallel dynamics to be included consistently. In isolated filament simulations in the TORPEX device \cite{HalpernPoP2014} using the GBS code \cite{RicciPPCF2012} filament dynamics were assumed to depend only on pressure, rather than on temperature and density separately. This is motivated by the observation that the dominant drive term in the current continuity equation, $\nabla\cdot\textbf{J} = 0$ is the divergence of the diamagnetic current, which is only pressure dependant. Other contributions arising from the polarization and parallel currents are not constant in pressure however, and deviations away from this have not been studied in 3D. It is important to characterize these effects in order to interpret the physics of filaments in both more complex fully turbulent simulations and in comparison with experiment.
\\This paper characterises the role of thermal effects on the dynamics of isolated filaments in three-dimensions using drift-reduced electrostatic two-fluid simulations carried out in the STORM module of BOUT++. The model used is an extension of that presented by Easy \emph{et.al} \cite{EasyPoP2014} with the inclusion of an evolution equation for the electron temperature. To systematically investigate the role of thermal effects the drive for the filament will be fixed across the majority of simulations by holding the pressure in the filament constant. The temperature and density in the filament will then be varied whilst ensuring that the pressure remains constant,  thereby allowing for the prominence of thermal effects to be systematically increased. The paper is organised as follows: Section 2 describes the model employed for simulations; section 3 describes modifications to the scaling of the characteristic filament velocity arising from the inclusion of thermal effects; section 4 presents results from the simulation scans at constant pressure, but with varying density and temperature amplitudes of the filament; section 5 discusses the results presented before section 6 concludes. A description of background profiles used in the simulations is given in appendix A.

\section{Model}
The model employed in this paper is an extension of that used by Easy \emph{et.al} \cite{EasyPoP2014} with the inclusion of a dynamic electron temperature. The equations of the model given here in normalized form, consist of density conservation
\begin{equation}
\label{Eqn:Dens}
\frac{\partial n}{\partial t} = -v\nabla_{||}n - \textbf{v}_{E}\cdot\nabla n - n\nabla_{||}v + g\left(\frac{\partial}{\partial y}nT - n\frac{\partial \phi}{\partial y}\right) + \mu_{n}\nabla_{\perp}^{2}n + S_{n}
\end{equation}
the vorticity equation (arising from current continuity)
\begin{equation}
\label{Eqn:Vort}
\frac{\partial \Omega}{\partial t} = -u\nabla_{||}\Omega -\textbf{v}_{E}\cdot\nabla_{\perp}\Omega + \frac{g}{n}\frac{\partial}{\partial y}nT + \nabla_{||}\left(u - v\right) + \frac{\left(u - v\right)}{n}\nabla_{||}n + \mu_{\Omega}\nabla_{\perp}\Omega
\end{equation}
parallel ion momentum
\begin{equation}
\frac{\partial u}{\partial t} = -u\nabla_{||}u - \textbf{v}_{E}\cdot\nabla u - \nabla_{||}\phi - \eta_{||}n\left(u-v\right) + 0.71\nabla_{||}T - \frac{uS_{n}}{n}
\end{equation}
parallel electron momentum
\begin{equation}
\frac{\partial v}{\partial t} = -v\nabla_{||}v - \textbf{v}_{E}\cdot\nabla v + m^{*}\left(\nabla_{||}\phi + \eta_{||}n\left(u-v\right) - 0.71\nabla_{||}T  + \frac{\nabla_{||}nT}{n}\right)
\end{equation}
and the electron energy equation
\begin{equation}
\label{Eqn:Temp}
\begin{split}
\frac{\partial T}{\partial t} = &-v\nabla_{||}T -\textbf{v}_{E}\cdot\nabla T \\ &+ 
\frac{2}{3n}\left(0.71n\left(u-v\right)\nabla_{||}T - \nabla_{||}q_{||}  - nT\nabla_{||}v - \eta_{||}n^{2}\left(u-v\right)^{2}\right) \\ &+ 
 \frac{2g}{3n}\left(T^{2}\frac{\partial n}{\partial y} - nT\frac{\partial \phi}{\partial y} + \frac{7}{2}nT\frac{\partial T}{\partial y} + \frac{1}{m^{*}}v^{2}\frac{\partial nT}{\partial y}\right) \\ &+ 
\frac{2}{3n}S_{E} + \frac{S_{n}v^{2}}{3m^{*}n} - \frac{TS_{n}}{n} + \mu_{T}\nabla_{\perp}^{2}T
\end{split}
\end{equation}
where $m^{*}$ is the mass ratio $m^{*} = m_{i}/m_{e}$. The state variables are the electron density $n$ (ion density inferred through quasineutrality), the parallel vorticity, $\Omega$, the electron temperature $T$, the electron parallel velocity $v$ and the ion parallel velocity $u$. The electrostatic potential, $\phi$ and the parallel conductive electron heat flux $q_{||}$ are given by the auxilliary equations
\begin{equation}
\Omega = \nabla_{\perp}^{2}\phi
\end{equation}
and 
\begin{equation}
q_{||} = -\frac{2}{7}\kappa_{||}\nabla_{||}T^{7/2} - 0.71nT\left(u-v\right)
\end{equation}
which is the Braginskii parallel heat flux \cite{Braginskii}. Collisional coefficients are the normalized resistivity
\begin{equation}
\eta_{||} = T^{-3/2}\frac{\nu_{ei}}{1.96\Omega_{e}}
\end{equation}
and the normalized parallel heat conductivity
\begin{equation}
\kappa_{||} = 3.16 \frac{T_{0}}{\nu_{ei}m_{e}c_{s}\rho_{s}}
\end{equation}
where the collision frequency is
\begin{equation} 
\nu_{ei} = \frac{n_{0}Z^{2}e^{4}\Lambda}{3\epsilon_{0}^{2}\sqrt{m_{e}\left(2\pi T_{0}\right)^{3}}}
\end{equation}
where $Z$ is the atomic number, $e$ is the elementary charge and $\Lambda$ is the Coulomb logarithm.
Perpendicular diffusion coefficients are \cite{FundamenskiNF2007}
\begin{equation}
\mu_{n} = \left(1 + 1.3q^{2}\right)\frac{\rho_{e}^{2}\nu_{e}}{\rho_{s}^{2}\Omega_{i}}
\end{equation}
\begin{equation}
\mu_{\Omega} = \frac{6}{8}\left(1 + 1.6q^{2}\right)\frac{\rho_{i}^{2}\nu_{i}}{\rho_{s}^{2}\Omega_{i}}
\end{equation}
\begin{equation}
\mu_{T} = 4.66\left(1 + 1.6q^{2}\right)\frac{\rho_{e}^{2}\nu_{e}}{\rho_{s}^{2}\Omega_{i}}
\end{equation}
 $\textbf{v}_{E}$ is the $\textbf{E}\times\textbf{B}$ drift velocity given by
\begin{equation}
\textbf{v}_{E} = \textbf{b}\times\nabla\phi
\end{equation}
$g$ represents the normalized curvature drive and is given by
\begin{equation}
g = \frac{2\rho_{s}}{R_{c}}
\end{equation}
where $R_{c}$ is the radius of curvature. $\Omega_{e} = eB/m_{e}$ and $\Omega_{i} = eB/m_{i}$ are the electron and ion gyro-frequencies respectively, $c_{s} = \sqrt{T_{e}/m_{i}}$ is the Bohm velocity and $\rho_{s} = c_{s}/\Omega_{i}$ is the Bohm gyro-radius. Finally $S_{n}$ and $S_{E}$ are particle and energy sources. The system of equations \ref{Eqn:Dens} to \ref{Eqn:Temp} have been normalized in the following manner 
\begin{equation}
\begin{array}{c c c c c}
n \rightarrow n_{0}n & &
u \rightarrow c_{s}u & &
v \rightarrow c_{s}v \\
\phi \rightarrow T_{0}\phi & &
T \rightarrow T_{0}T & &
t \rightarrow t/\Omega_{i} \\
L \rightarrow \rho_{s}L & &
S_{n} \rightarrow n_{0}\Omega_{i}S_{n} & &
S_{E} \rightarrow n_{0}T_{0}\Omega_{i}S_{E}
\end{array}
\end{equation}
where $L$ indicates length parameters and $t$ is time. For the simulations detailed within this paper parameters relevant to a MAST Ohmic L-mode are used, given below in table \ref{Tbl:params}.
\begin{table}[htbp]
\begin{tabular}{l || l || l}
Parameter & Description & Value \\
\hline
\multicolumn{3}{l}{Input parameters} \\
\hline
$n_{0}$ & Reference density & $0.8\times 10^{19}$m$^{-3}$ \\
$T_{0}$ & Reference temperature & $40$eV \\
$B$ & Magnetic field strength & $0.5$T \\
$R_{c}$ & Curvature radius & $1.5$m \\
$L_{||}$ & Parallel connection length & $10$m \\
\hline
\multicolumn{3}{l}{Derived parameters} \\
\hline
$\rho_{s}$ & Bohm gyro-radius & $1.82$mm \\
$c_{s}$ & Sound speed & $43.9$km/s \\
$\Omega_{i}$ & Ion gyro-frequency & $24.1$MHz \\
$\Omega_{e}$ & Electron gyro-frequency & $87.9$GHz \\
$\nu_{ei}$ & Collision frequency & $1.22$MHz \\
$\eta_{||}$ & Normalized resistivity & $7.08\times 10^{-6}$ \\
$\kappa_{||}$ & Normalized heat conductivity & $2.28\times 10^{5}$ \\
$g$ & Normalized curvature drive & $2.43\times 10^{-3}$
\end{tabular}
\caption{Table of input and derived parameters used for simulations presented within this paper. Values are taken to be representative of the SOL at the outboad midplane of a MAST double null plasma.}
\label{Tbl:params}
\end{table}
\\The simulation geometry is a 3D slab with $x$ and $y$ representing the drift plane perpendicular to the magnetic field and $z$ as the direction along the magnetic field line. $x,y$ and $z$ are a Cartesian coordinate system in which $x$ representing the direction normal to both the magnetic field line and magnetic flux surfaces and $y$ representing the bi-normal direction which is normal to the magnetic field line but lies within a flux surface. This simulation geometry is identical to that of Easy \emph{et.al} \cite{EasyPoP2014}. Magnetic curvature is present with a curvature vector $\bm{\kappa} = -\hat{\textbf{x}}/R_{c}$ such that $\textbf{b}\times\bm{\kappa}\cdot\nabla = -\frac{1}{R_{c}}\frac{\partial}{\partial z}$ which leads to the form of the parameter $g$. The factor of $2$ in $g$ accounts for the curvature and $\nabla B$ contributions. The domain is assumed to be symmetric about $z=0$ with sheath boundary conditions at $z = \pm L_{||}$ with the half of the domain between $z=0$ and $z=L_{||}$ simulated. At the midplane boundary in $y$ symmetry boundary conditions are applied which impose Neumann boundary conditions on scalar variables ($n$, $T$, and $\Omega$) and Dirichlet boundary conditions on flux variables ($u$, $v$ and $q_{||}$). At the target boundary the Bohm sheath boundary conditions are applied to the ion and electron velocities such that
\begin{equation}
u|_{z=L_{||}} = \sqrt{T}
\end{equation}
\begin{equation}
\label{Eqn:vsheath}
v|_{z=L_{||}} = \sqrt{T}\exp\left(-\left(V_{f} + \frac{\phi}{T}\right)\right)
\end{equation}
where $V_{f}$ is the normalized floating potential 
\begin{equation}
V_{f} = \frac{1}{2}\ln\left(\frac{2\pi m_{e}}{m_{i}}\right) 
\end{equation}
and the wall potential is taken to be zero. The parallel heat flux is set by the sheath energy flux density \cite{Stangeby}
\begin{equation}
\label{Eqn:Qsheath}
Q_{||}|_{z=L_{||}} = \gamma nT^{3/2}
\end{equation}
where $\gamma \approx 2 - V_{f} = 5.21$ is the electron sheath transmission coefficient. The parallel energy flux density is related to the parallel heat flux density by 
\begin{equation}
\label{Eqn:Qpar}
Q_{||} = \frac{5}{2}nTv + \frac{1}{2}nm_{e}v^{3} + q_{||}
\end{equation}
so that, by substitution of equation (\ref{Eqn:vsheath}) into equation (\ref{Eqn:Qpar}) and then by substitution into equation (\ref{Eqn:Qsheath}) an expression for the sheath heat flux is given by
\begin{equation}
q_{||}|_{z=L_{||}} = nT^{3/2}\left[\gamma - \frac{5}{2}\exp\left(-\left(V_{f} + \frac{\phi}{T}\right)\right) - \frac{1}{2}m^{*}\exp\left(-3\left(V_{f} + \frac{\phi}{T}\right)\right)\right]
\end{equation}
\\ Filaments are initialized as a Gaussian in the drift plane whilst being either homogeneous or having a tanh function in the parallel direction. Filaments are seeded on top of a pre-determined plasma background. In this paper the plasma background is constant for all simulations and is described in Appendix A. The simulation mesh contains 128 grid points in $x$ and $y$ and 64 grid points in $z$. The size of the $x$ and $y$ domain, $L_{x}$ and $L_{y}$ is $10\delta_{\perp}\times 10\delta_{\perp}$ where $\delta_{\perp}$ is the Gaussian width of the filament at initialization. The size of the grid in the $z$ direction is $L_{z} = L_{||} = 10$m.
\\There are a number of assumptions that have been made in the setup of this model. Equations \ref{Eqn:Dens} to \ref{Eqn:Temp} are derived assuming a drift-ordering, neglecting electromagnetic effects and assuming cold ions. In addition the Boussinesq approximation is made to simplify the form of the polarization current term in the vorticity equation such that $\nabla\cdot \left(n\nabla_{\perp}\phi\right) \rightarrow n\nabla_{\perp}^{2}\phi$. The term $\frac{2gv^{2}}{3nm^{*}}\frac{\partial nT}{\partial z}$ in equation (\ref{Eqn:Temp}) arises from the inclusion of the electron gyroviscosity, which is required as an energy transfer channel from the gyroviscous cancellation in the parallel electron momentum equation. The system does not fully conserve energy for two reasons: 1. Advection due to the ion polarization velocity is neglected in the ion parallel momentum equation which introduces an energy sink into the parallel ion kinetic energy. This problem is present in most drift-ordered fluid models, however correcting this requires the code to advect with the ion polarization velocity, which is not currently possible within BOUT++. 2. The Boussinesq approximation introduces an energy sink in the perpendicular $\textbf{E}\times\textbf{B}$ kinetic energy and has recently been studied in the context of filaments \cite{AngusPoP2014}. This can have a strong impact on large density amplitude filaments, however for amplitudes studied in this paper the impact is likely to be minimal. This can be corrected by relaxing the Boussinesq approximation which may be achieved using a new multi-grid Laplacian inversion algorithm in BOUT++ \cite{OmotaniPPCF2015}, the details of which will be reported at a later date. Unfortunately with the present model, which contains fast times scales from electron heat dynamics, the use of the multi-grid solver becomes computationally inhibitive and requires significant optimization. For this reason the simulations carried out here continue to make the Boussinesq approximation. Nevertheless the energy sinks that arise from these two processes are at least one order (in terms of the magnetization parameter $\delta$) below the dominant energy transfer channels in filament dynamics and are not expected to have a significant impact on the results presented herein. It is also worth noting that isothermal models of filaments inherently break energy conservation through neglect of Joule heating. Once again this is a weak effect, but is rectified here in the electron temperature evolution.

\section{Velocity Scaling}
\begin{figure}[htbp]
\centering
\includegraphics[width=0.5\textwidth]{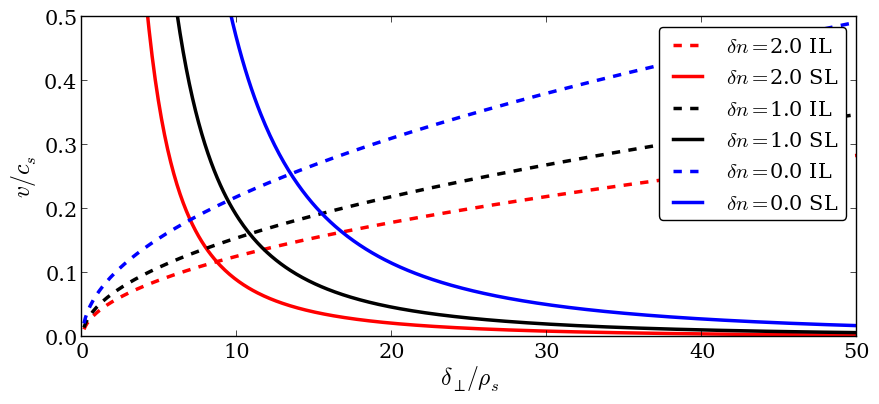}
\caption{Scaling of the 2D filament characteristic velocity in the inertial regime (IL, broken lines) calculated from equation (\ref{Eqn:vI}) and in the sheath limited regime (SL, solid lines) calculated from equation (\ref{Eqn:vsh}) for the following cases: $\delta n  = 2, \delta T = 0$ (red), $\delta n  = 1, \delta T = 1$ (black) and $\delta n  = 0, \delta T = 2$ (blue). }
\label{Fig:velocity_scaling}
\end{figure}

Whilst in general the dynamics of filaments are complex and non-linear, simple scaling laws for a characteristic radial filament velocity have been shown to capture the non-linear motion observed in simulations relatively well in both 2D and 3D \cite{EasyPoP2014,WalkdenNF2015}. The inclusion of thermal effects modifies both drive and dissipation terms in the vorticity equation so has the potential to modify the scaling of the characteristic radial filament velocity. To derive a scaling law for the characteristic radial velocity the vorticity equation (equation \ref{Eqn:Vort}) it is first decoupled into two equations describing the evolution of the odd and even parity (in the bi-normal direction about the filament centre) components of the electrostatic potential, $\phi^{o}$ and $\phi^{e}$ respectively. Note that through $\textbf{E}\times\textbf{B}$ motion $\phi^{o}$ produces radial motion of the filament whilst $\phi^{e}$ produces circulatory motion about the filament centre. The decoupling technique is introduced and discussed in ref \cite{WalkdenNF2015} and gives
\begin{equation}
\label{Eqn:vort+}
\frac{\partial \Omega^{e}}{\partial t} + \textbf{v}_{E}^{o}\cdot\nabla\Omega^{o} + \textbf{v}_{E}^{e}\cdot\nabla\Omega^{e} = \frac{1}{n}\nabla_{||}J_{||}^{e}
\end{equation}
\begin{equation}
\frac{\partial \Omega^{o}}{\partial t} + \textbf{v}_{E}^{o}\cdot\nabla\Omega^{e} + \textbf{v}_{E}^{e}\cdot\nabla\Omega^{o} = \frac{g}{n}\frac{\partial nT}{\partial y} + \frac{1}{n}\nabla_{||}J_{||}^{o}
\end{equation}
where
\begin{equation}
\textbf{v}_{E}^{e} = \textbf{b}\times\nabla\phi^{o}
\end{equation}
\begin{equation}
\textbf{v}_{E}^{o} = \textbf{b}\times\nabla\phi^{e}
\end{equation}
If parallel variation is weak (as is the case here where filaments are homogeneous in the parallel direction) the above equations can be integrated along the magnetic field line which, along side linearisation of the sheath boundary conditions, gives
\begin{equation}
\label{Eqn:vort+2D}
\frac{\partial \Omega^{e}}{\partial t} + \textbf{v}_{E}^{o}\cdot\nabla\Omega^{o} + \textbf{v}_{E}^{e}\cdot\nabla\Omega^{e} = \frac{1}{L_{||}\sqrt{T}}\left(TV_{f} + \phi^{e}\right)
\end{equation}
\begin{equation}
\label{Eqn:vort-2D}
\frac{\partial \Omega^{o}}{\partial t} + \textbf{v}_{E}^{e}\cdot\nabla\Omega^{o} + \textbf{v}_{E}^{o}\cdot\nabla\Omega^{e} = \frac{g}{n}\frac{\partial nT}{\partial y} + \frac{\phi^{o}}{L_{||}\sqrt{T}}
\end{equation}
In this form the floating potential appears explicitly in equation (\ref{Eqn:vort+2D}) and drives the growth of $\Omega^{e}$ and consequently $\phi^{e}$ thus producing a source of circulatory motion. This is analogous however to the isothermal 3D case where the Boltzmann response produces finite $\phi^{e}$ thereby leading to circulatory motion \cite{WalkdenNF2015}. Note that in the isothermal 2D limit $V_{f}T_{e,0}$ can be simply treated as a gauge for $\phi^{e}$. An estimate for the velocity of the filament radially away from its centre can be determined by balancing drive and dissipation terms in  (\ref{Eqn:vort-2D}). At present the non-linear coupling between $\phi^{e}$ and $\phi^{o}$ is assumed small as the initial filament motion develops. This term represents the charge-mixing employed by Myra \emph{et.al} in the case of 2D spinning blobs, however this does not strongly impact filaments studied here, as will be shown in a later section. For now this term is assumed to be small with justification being provided \emph{a posteriori}. This gives an expression for the potential, $\phi^{o}$, which can be converted into a velocity by
\begin{equation}
v_{R} = v = \left| \textbf{b}\times\nabla\phi^{o}\right| \sim \phi^{o}/\delta_{\perp}
\end{equation}
The inertial regime occurs when the inertial term balance the drive\cite{GarciaPoP2006}. Taking estimates for gradients as $\nabla_{\perp},\partial/\partial z \rightarrow 1/\delta_{\perp}$ this leads to 
\begin{equation}
\label{Eqn:vI}
v_{I} \sim \sqrt{\delta_{\perp}g\frac{\delta p}{n_{0} + \delta n}}
\end{equation}
The sheath regime occurs when the drive is balanced by the sheath dissipation term \cite{KrashenninikovPLA2001} and gives a velocity
\begin{equation}
\label{Eqn:vsh}
v_{Sh} \sim \frac{gL_{||}\delta p \sqrt{T_{0} + \delta T}}{\left(n_{0} + \delta n\right)\delta_{\perp}^{2}}
\end{equation}
In the above two expressions the density and temperature field have been decomposed into backgrounds, $T_{0},n_{0}$ and perturbations $\delta n, \delta T$ with the pressure perturbation defined as $\delta p = T_{0}\delta n + n_{0}\delta T + \delta n\delta T$. As noted by Omotani \emph{et al} \cite{OmotaniPPCF2015} coefficients of order unity should in principle be included prior to these factors, however this is unnecessary for the basic derivation carried out here. The fundamental filament size \cite{YuPoP2003,MyraPoP2006}, $\delta^{*}$, is the size at which the two regimes exhibit the same velocity such that 
\begin{equation}
v_{I} \sim v_{Sh} \Rightarrow \delta^{*} = \left(gL_{||}^{2}\delta p \frac{T_{0} + \delta T}{n_{0} + \delta n}\right)^{1/5}
\end{equation}
Thermal effects change both the magnitude of the characteristic velocity and $\delta^{*}$ which characterises the transition from one regime to another. Figure \ref{Fig:velocity_scaling} shows the characteristic velocity scaling in the inertial (broken lines) and sheath limited (solid lines) regimes for three different cases where pressure is held constant, but $\delta n$ and $\delta T$ are varied in the ranges $\delta n = [2,1,0]$ and $\delta T = [0,1,2]$.
The velocities in each regime, alongside the fundamental filament size vary as the temperature and density in the filament change (keeping pressure fixed) with a general increase as the temperature is increased and the density decreased. 
\\Angus \emph{et al} \cite{AngusPoP2014} and Omotani \emph{et al} \cite{OmotaniPPCF2015} have recently shown that, with the Boussinesq assumption relaxed the inertial regime velocity scaling is better represented by taking the background inertia ($n_{0}$) rather than the full filament inertia ($n_{0} + \delta n$) in the inertial term. This has the effect of collapsing the three inertial velocity curves onto the single curve with $\delta n = 0$ such that, whilst the velocity in the inertial regime now becomes constant for constant pressure, the change to the fundamental filament size with $\delta T$ remains similar to the Boussinesq case. For the rest of this paper, when the fundamental filament size is referred to, the value with $\delta n = \delta T$ will be given. The value of $\delta^{*}$ varies by a factor of order unity which, given that the theory presented above is only accurate to factors of order unity, may be considered a relatively small variation. The theory outlined here predicts a net increase in the characteristic filament velocity when the pressure in the filament is contained predominantly in the temperature. Characteristic velocity analysis is useful to identify regimes of filament dynamics, but fails to capture fully the evolution of the filament. To fully study the role of thermal effects 3D simulations have been conducted.  

\section{Simulation Results}
To assess the role of thermal effects on filament dynamics a set of simulations has been conducted varying the amplitude of the filament at initialization in both density, $\delta n$, and temperature, $\delta T$. To isolate effects associated with the introduction of temperature variation into the model the total pressure of the filament at initialization is held constant such that 
\begin{equation}
\left(n_{eq} + \delta n\right)\left(T_{eq} + \delta T \right) \approx 3
\end{equation}
which, with $T_{eq} = 1.28$ and $n_{eq} = 0.78$ taken from figure \ref{Fig:eq_comp} in normalized units (see table \ref{Tbl:params}), leads to the following three combinations of $\delta n$ and $\delta T$ used for simulations
\begin{equation}
\label{Eqn:amplitudes}
\begin{array}{c c c c c}
\delta n = 1.59 & & \delta n = 0.73 & & \delta n = 0 \\
\delta T = 0    & & \delta T = 0.73 & & \delta T = 2.61
\end{array}
\end{equation}
These simulations keep the diamagnetic current at initialization constant despite the variation in the amplitudes. When $\delta_{\perp} \sim \delta^{*}$ the drive for the the filament motion is balanced by a combination of inertial and sheath dissipation. It also represents filament sizes commonly found in MAST \cite{DudsonPPCF2008}, for which the simulation parameters used here are representative. Since filaments with $\delta_{\perp} \sim \delta^{*}$ are likely to propagate furthest into the SOL it is sensible to first study the role of thermal effects in these conditions.
\subsection{Constant pressure scan}
\begin{figure}[htbp]
\centering
\includegraphics[width=0.85\textwidth]{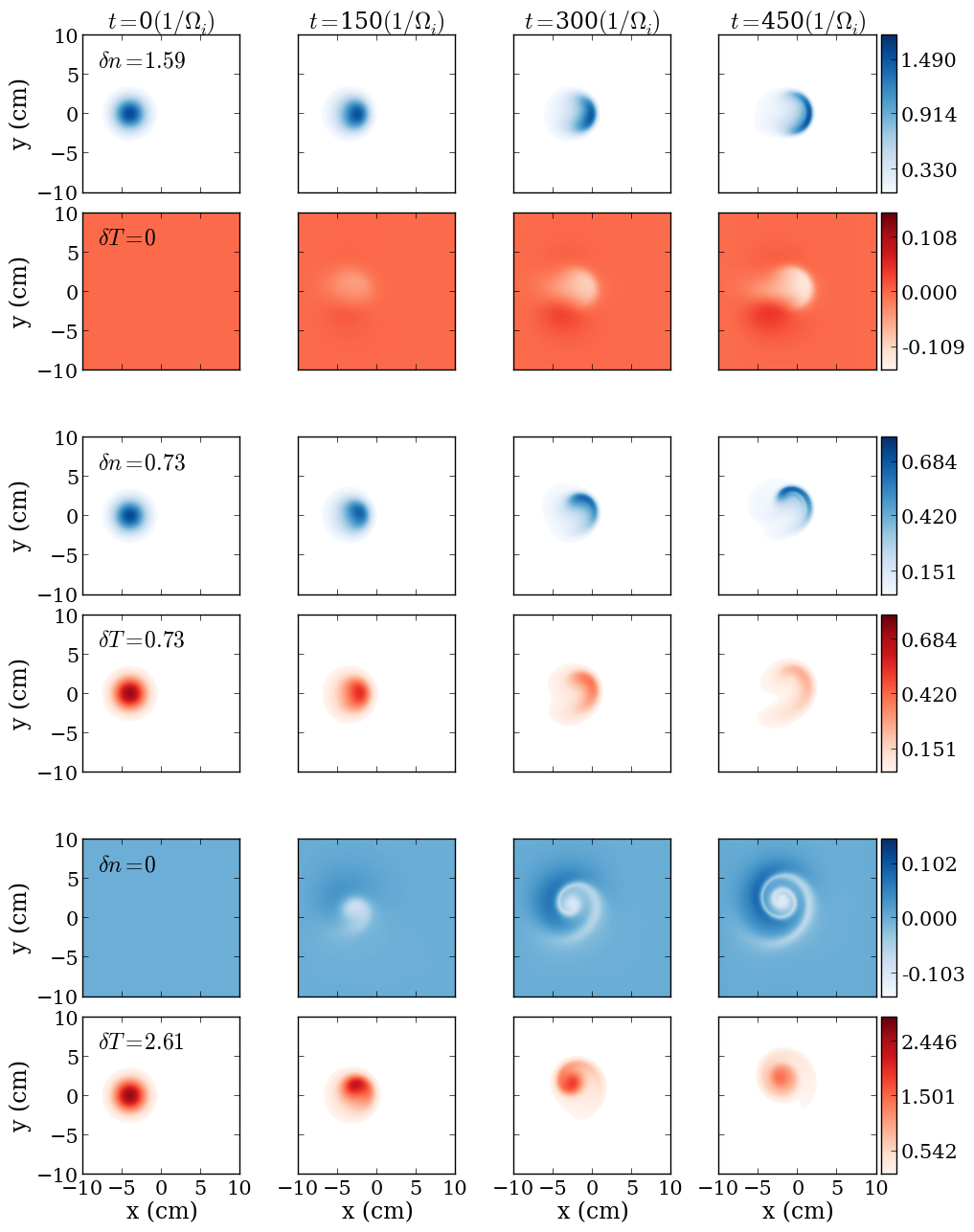}
\caption{Cross sections of density (blue color scales) and temperature (red color scales) at the midplane shown at different points in the filament evolution. The three simulations corresponding to the three pairings of $\delta n$ and $\delta T$ given in (\ref{Eqn:amplitudes}) are shown.}
\label{Fig:Inertial_Xsecs}
\end{figure}

\begin{figure}[htbp]
\centering
\includegraphics[width=0.7\textwidth]{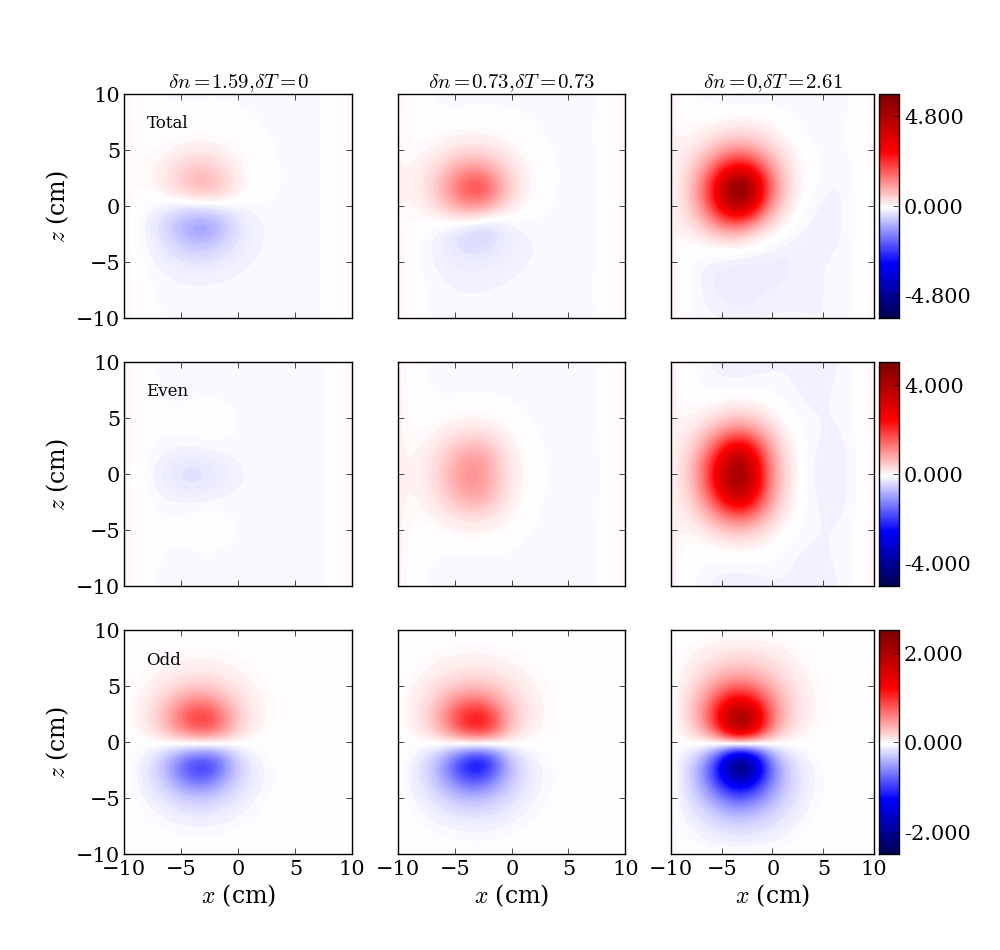}
\caption{Electrostatic potential, $\phi$ (upper row) and its even and odd components (middle and lower rows respectively) about the position of the filament centre in the bi-normal direction for each pair of amplitudes $\delta n$ and $\delta T$. Profiles are shown at the midplane at $t = 150/\Omega_{i}$ which is approximately at the maximum on the filament radial velocity.}
\label{Fig:Inertial_sym_components}
\end{figure}

\begin{figure}[htbp]
\centering
\includegraphics[width=0.5\textwidth]{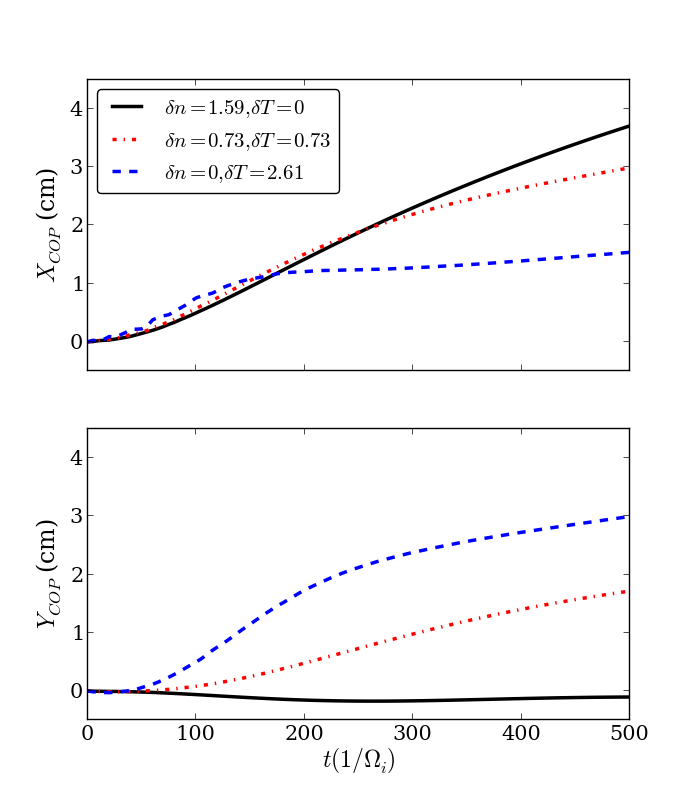}
\caption{Centre-of-pressure position in the radial direction ($x$) (upper) and in the bi-normal direction ($z$) (lower) of the filament cross-section at the midplane in each of the simulations corresponding to the pairings of $\delta n$ and $\delta T$ given in (\ref{Eqn:amplitudes}).}
\label{Fig:Inertial_COM_const_p}
\end{figure}

\begin{figure}[htbp]
\centering
\includegraphics[width=\textwidth]{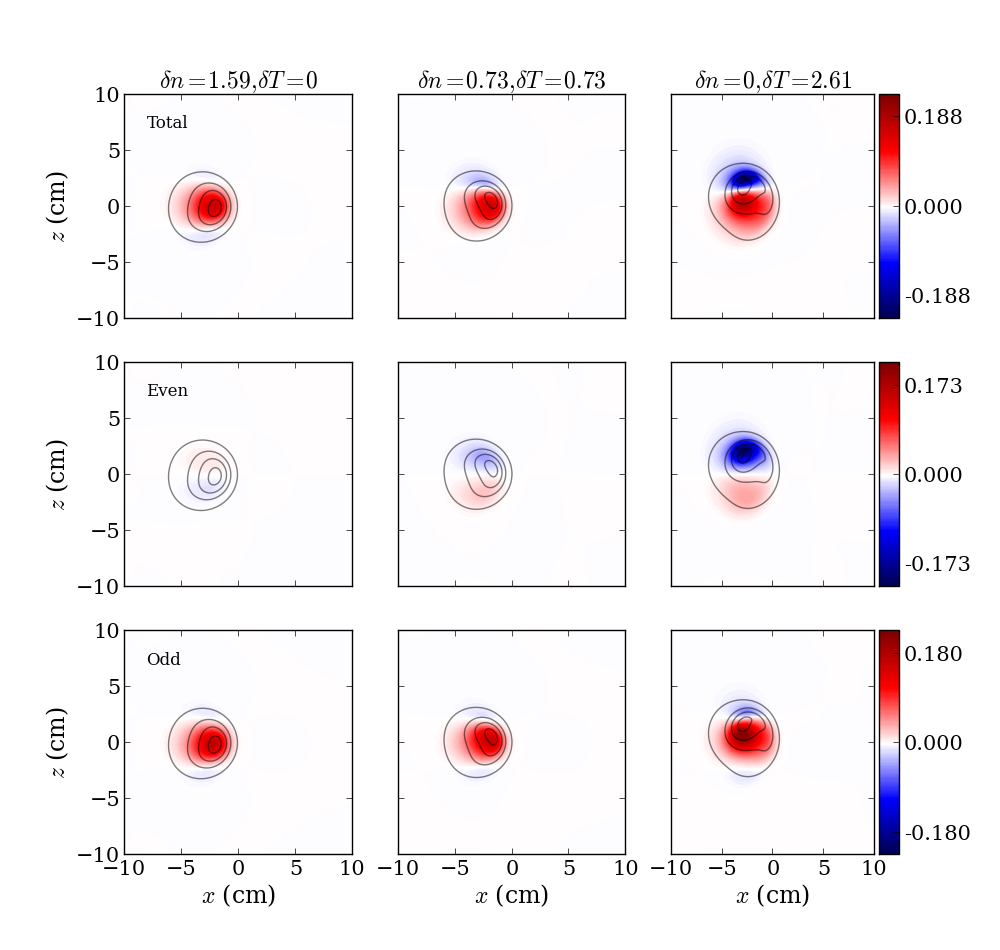}
\caption{Radial $\textbf{E}\times\textbf{B}$ heat flux from the total (upper), even (middle) and odd (lower) component of the potential at $t=150/\Omega_{i}$}
\label{Fig:ExB_flux_d2}
\end{figure}

\begin{figure}[htbp]
\centering
\includegraphics[width=0.6\textwidth]{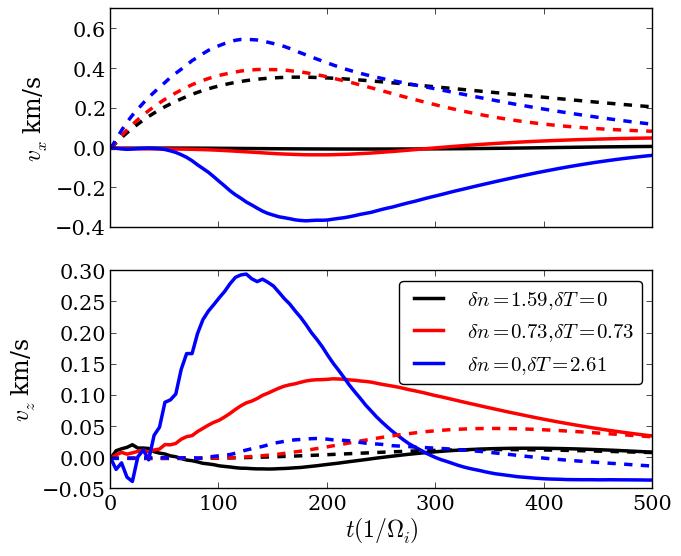}
\caption{Cross-section averaged radial (upper) and bi-normal (lower) average $\textbf{E}\times\textbf{B}$ velocity for the three cases in the constant pressure scan. Shown are contributions from the even (solid lines) and odd (broken lines) components of the potential.}
\label{Fig:d2_ExB_const_p_sym}
\end{figure}

\begin{figure}[htbp]
\centering
\includegraphics[width=0.4\textwidth]{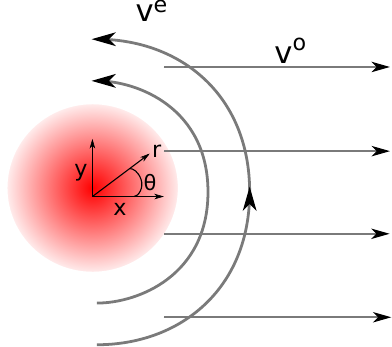}
\caption{Schematic illustrating the geometry of the filament flow pattern with a constant flow in $x$ of magnitude $v^{-}$ and a circulatory flow in the positive $\theta$ direction of magnitude $v^{+}$.}
\label{Fig:vel_schematic}
\end{figure}

\begin{figure}
\centering
\includegraphics[width=0.5\textwidth]{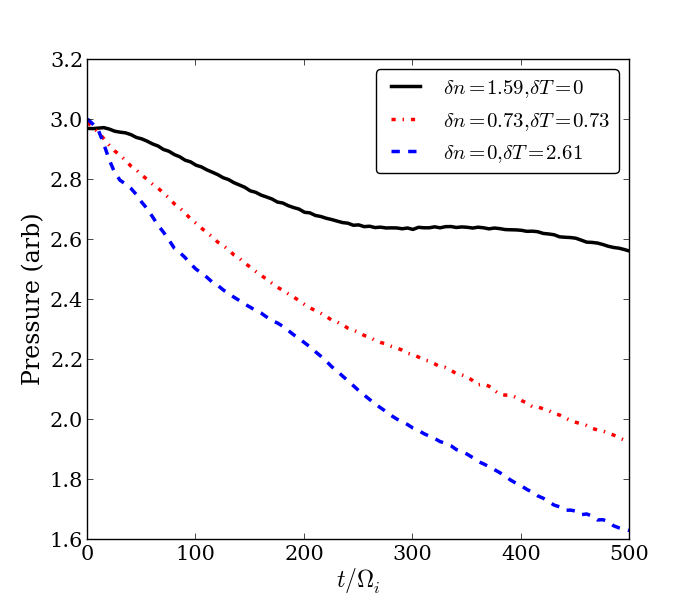}
\caption{Maximum pressure in the filament at the midplane as a function of time for each combination of $\delta n, \delta T$}
\label{Fig:Pressure_loss}
\end{figure}

Figure \ref{Fig:Inertial_Xsecs} shows the evolution of the filament cross-section at the midplane for each of the three combinations of $\delta n$ and $\delta T$, with a cross-field width of 2cm and a homogeneous parallel profile.\\ 
There are clear differences in the cross-field dynamics of the filament as the distribution of the filament pressure at initialization is varied between the temperature and density. The main differences are: \begin{itemize}
\item The filament exhibits a strong increase in its bi-normal propagation and remains more coherent with a suppression of the 'mushrooming' as $\delta T$ increases (and correspondingly $\delta n$ decreases).
\item  As the coherence and bi-normal propagation increase, the radial propagation of the filament is suppressed .
\end{itemize}
Before these features are analysed in detail however, is worth commenting on the structure that forms in the density and/or temperature fields when no initial filament amplitude is present. In each of the two cases (ie when either $\delta n = 0$ or when $\delta T = 0$) the structure that appears in the initially empty field is a result of the dependency of the boundary conditions at the sheath. If there is an initial density perturbation but no temperature perturbation then the sheath heat flux, which depends linearly on the density, is locally enhanced in the filament and the sink acting on the temperature increases. This produces a local cooling of the filament below the background. Compression of magnetization and $\textbf{E}\times\textbf{B}$ drifts then acts on the perturbation to produce the structure observed in the simulations. The analogous effect occurs with $\delta n = 0$ due to a local enhancement of the sheath velocity increasing the sink of density. The origin of these structures is therefore physically consistent, however they have minimal impact on the filament dynamics since they are small perturbations compared to the comparatively large amplitudes of the filaments themselves.
\\The increased propagation of the filament in the bi-normal direction and the reduced propagation in the radial direction are both a result of the temperature dependence of the floating potential, which appears as a drive term in equation (\ref{Eqn:vort+2D}), and leads to growth of the even component of the potential, $\phi^{e}$. The even component of the potential provides a circulatory motion which in itself cannot provide net propagation in either the radial or the bi-normal direction. However when superposed with the radial motion induced by the odd component of potential the filament exhibits propagation in both the radial and bi-normal direction. This is similar to the observation of filament motion in the bi-normal direction due to the Boltzmann response \cite{AngusPoP2012,WalkdenPPCF2013,EasyPoP2014,WalkdenNF2015} but occurs in the absence of parallel gradients. There is no clear sign of the rotational instability observed in ref \cite{MyraPoP2004} which may be a result of the use of physically motivated collisional dissipation in this study. Figure \ref{Fig:Inertial_sym_components} shows the total, even and odd components of the potential for each simulation in the scan of $\delta n,\delta T$ combinations, taken at $t = 150/\Omega_{i}$. At this point $\phi^{e}$ and $\phi^{o}$ have developed and go on to determine the evolution of the filament. 
\\As predicted by equation (\ref{Eqn:vort+2D}), the even component of the potential increases as the temperature contribution to the filament pressure becomes large with the total potential becoming dominated by the even component when $\delta n =0, \delta T = 2.61$. Interestingly the odd component of the potential also increases. This is notable since it indicates that charge mixing, as proposed in ref \cite{MyraPoP2004}, is not suppressing the odd component of the potential in this case. Instead the increase occurs as is predicted by the characteristic radial velocity of the filament obtained from the simplified 2D calculations in section 3. This justifies neglect of the charge mixing terms in section 3. The closure channels in the current circuit arising from both cross-field polarization currents (LHS of equation (\ref{Eqn:vort-2D})) and sheath currents (second RHS term of equation (\ref{Eqn:vort-2D})) are not constant at constant pressure; the polarization current has a dependency on density but not on temperature whilst the sheath current has a dependency on $nT^{-1/2}$. In both cases, when $\delta T/T_{0} >> \delta n/n_{0}$ the effective resistance of the closure channel is increased and the filament can develop a larger potential which leads to a larger odd component of the potential and therefore a larger radial velocity. As noted in section 3 the Boussinesq approximation can affect these results with non-Boussinesq simulations showing that the inertial dissipation may be better represented by the background than the filament density. This has no affect on the sheath dissipation however, so the result that the odd component of the potential increases is still applicable. It was not possible to conduct non-Boussinesq simulations due to the greatly increased computational demand that they entail, however investigation of this in a thermal model should be pursued in the future. 
\\Whilst the increase in the odd component of the potential can be explained by 2D scalings, it is in contrast to the net decrease in radial propagation observed in the filament.  This has been quantified by taking the centre-of-pressure (COP, used instead of the centre of mass to account for the case when $\delta n = 0$) of the filament cross-section. Figure \ref{Fig:Inertial_COM_const_p} shows the position of the filament COP, calculated at the midplane, for each of the combinations of $\delta n$ and $\delta T$. 
The filament COP in the binormal direction shows a significant increase as the pressure is carried by the temperature of the filament rather than the density. This occurs as the even component of the potential increases and rotates the filament leading to net propagation in the bi-normal direction. By contrast the propagation of the radial COP coordinate occurs in a similar manner for all three filaments in the early stages (approximately first third) of the simulation. There is a minor deviation in the case with $\delta n = 0$ which coincides with the increase in the odd component of the potential and agrees with the larger characteristic velocity predicted by the 2D scaling analysis, however this variation is weak compared to the differences in the net propagation of the filament where the three cases decouple. This reduction in the net propagation occurs as the filament begins to move radially, but gets entrained in the circulatory flow from the even component of the potential. To illustrate this the radial $\textbf{E}\times\textbf{B}$ heat flux, given by $p\textbf{v}_{E,x}$,
 has been calculated and is shown across the filament cross-section at the midplane at $t = 150/\Omega_{i}$ in figure \ref{Fig:ExB_flux_d2}.
\\The even component of potential in the $\delta n = 0$ case provides a re-circulation of the pressure which, when superposed with the radial outflux due to the odd component, leads to regions in the filament where the net radial motion is inwards as opposed to outwards. As shown by Garcia \emph{et.al} \cite{GarciaPoP2006}, averaging the $\textbf{E}\times\textbf{B}$ particle flux across the filament cross-section is equivalent to the centre-of-mass analysis, and by analogy here averaging the $\textbf{E}\times\textbf{B}$ heat flux is equivalent to the COP velocity. Figure \ref{Fig:d2_ExB_const_p_sym} shows the cross-section averaged $\textbf{E}\times\textbf{B}$ heat flux, defined by
\begin{equation}
\label{Eqn:v_COP}
v_{x} = \frac{\int\int nT\frac{\partial \phi}{\partial y}dxdy}{\int\int nT dx dy}
\end{equation}
\[ v_{y} = \frac{\int\int nT\frac{\partial \phi}{\partial x}dxdy}{\int\int nT dx dy}\]
in the radial and bi-normal directions for the even (shown as solid lines) and odd (broken lines) components of the potential.
\\Figure \ref{Fig:d2_ExB_const_p_sym} shows that the total radial velocity becomes significantly affected by the even component of the potential when $\delta n = 0$. This reduction induced by the even component can be elucidated by considering a simplified model where the total velocity field of the filament is the superposition of a velocity, $v^{o}$ arising from the odd component which is constant in magnitude and directed in the positive x direction, and a circulatory velocity arising from the even component, $v^{e}$. The net velocity in the radial direction is then
\begin{equation}
\label{Eqn:v_simplified}
v_{x} = v^{o} + v^{e}\textbf{e}_{x}\cdot\textbf{e}_{\theta} = v^{o} - v^{e}\sin \left(\theta\right)
\end{equation}
where $r,\theta$ define polar coordinates centred on the filament, with $r$ being the radial distance from the filament centre and $\theta$ being the angle subtended around the filament anti-clockwise, and $\textbf{e}_{x}$ is the unit vector in the x direction. Figure \ref{Fig:vel_schematic} outlines this setup schematically.
Substituting equation (\ref{Eqn:v_simplified}) into equation (\ref{Eqn:v_COP}) gives a COP velocity of
\begin{equation}
\langle v_{R}\rangle = v^{o} - v^{e}\frac{\int r nT\sin\theta dr d\theta}{\int rnTdrd\theta}
\end{equation}
If $ p = nT = p\left(r\right)$ then $\langle v_{x} \rangle = v^{o}$ and the even component of the potential has no impact on the radial motion. If, on the other hand, the pressure has an asymmetry in $\theta$ such that $p = p\left(r,\theta\right)$ then the second term above is not negligible. Indeed when the pressure is larger in the range $\theta = 0,\pi$ than in the range $\theta = \pi,2\pi$ then the net radial velocity is reduced by the circulatory motion of $v^{e}$. As a counter example a filament that maintains its symmetry in $\theta$ for the duration of its evolution cannot experience a reduction in its radial motion due to $v^{e}$. Returning to figure \ref{Fig:Inertial_Xsecs}, as the filament is advected by the flow formed by $v^{o}$ and $v^{e}$ it indeed forms a structure which provides the conditions to reduce the net radial COP velocity, as is observed in figure \ref{Fig:d2_ExB_const_p_sym}. In the full simulation the flow fields that develop are more complex than those used here to describe the effect of the reduction to the radial COP velocity, and furthermore the precise shape of the pressure is difficult to predict analytically. Nevertheless this simple analysis describes the features observed in the more complex simulations and provides an understanding of the cause of the reduction to the radial COP velocity. A good topic for development would be an analytical understanding of how the filaments symmetry is affected by its flow-field. 
\\As the filaments evolve there is a notable reduction in the filament pressure, which occurs at different rates when the pressure is lost through particle advection or heat conduction into the sheath. Figure \ref{Fig:Pressure_loss} shows the drop in the peak pressure of the filament at the midplane for the three cases for $\delta n$ and $\delta T$ studied in the previous section.
\\The loss of pressure occurring in the filament with $\delta n = 0$ occurs more rapidly than the others as the loss occurs predominantly through the sheath heat flux. This reduces the filament pressure and ultimately reduces the drive for the motion as the simulation continues.
\subsection{Increased filament pressure}
\begin{figure}[htbp]
\centering
\includegraphics[width=0.5\textwidth]{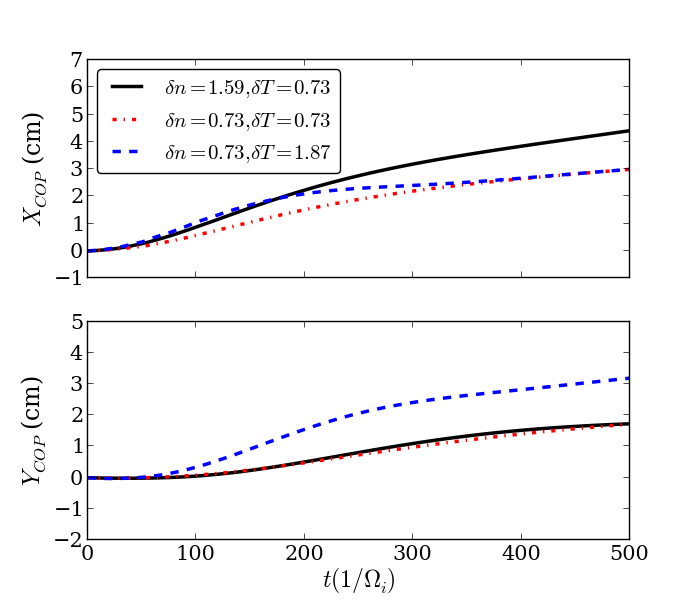}
\caption{Centre of pressure position in the radial direction ($x$) (upper) and in the bi-normal direction ($z$) (lower) of the filament cross-section at the midplane in each of the simulations corresponding to the pairings of $\delta n$ and $\delta T$ given in equation (\ref{Eqn:amplitudes_p_var}).}
\label{Fig:Inertial_COM_var_p}
\end{figure}

To test the effect of an increase in the pressure of the filament on its motion, a further two simulations have been run with the following amplitudes:
\begin{equation}
\label{Eqn:amplitudes_p_var}
\begin{array}{c c c c c}
\delta n = 1.59 & & \delta n = 0.73  \\
\delta T = 0.73    & & \delta T = 1.87 
\end{array}
\end{equation}
These simulations represent an increase in pressure over the previous simulations from $\delta p = 3$ to $\delta p = 4.75$, with the total increase held fixed between the two cases. The way in which the pressure is increased however has been varied between an increase in the density perturbation and an increase in the temperature perturbation. Figure \ref{Fig:Inertial_COM_var_p} compares the COP coordinates for the two filaments outlined above with the filament at $\delta n = 0.73, \delta T = 0.73$.
\\The radial COP propagation of the filament shows an increase with increasing pressure in the filament in the early stages before the even component of the potential plays a significant role, however once again the net propagation differs significantly between the two cases due to the filaments circulatory motion. This effect is extreme enough that, when the pressure in the filament is increased through the electron temperature channel alone, the net propagation of the filament is actually slightly reduced compared to the lower pressure case.  The bi-normal COP propagation on the other hand is dependant only on the temperature perturbation, and remains similar if $\delta T$ is held constant despite an increase in pressure. This suggests that care must be taken when treating either the temperature or density as a free parameter; if temperature is known but density is not then a change in density may result in a change to the radial motion of the filament though the bi-normal motion may remain fixed. If the opposite is true then a change in temperature can affect both the radial and bi-normal motion. This may also be seen as a distinguishing feature however. If, for example, 2D light emission is imaged (as in GPI \cite{FuchertPPCF2014} or BES \cite{MoultonJNM2015}) and two filaments are identified with similar amplitudes but vastly different motions then the signatures of their motions may be used to help constrain the possible combinations of density and temperature required for the observed light emission. This method has recently been used to distinguish filament dynamics on MAST \cite{MilitelloPrePrint2016} 
\subsection{Effect of parallel lengthscale}
\begin{figure}[htbp]
\centering
\includegraphics[width=0.8\textwidth]{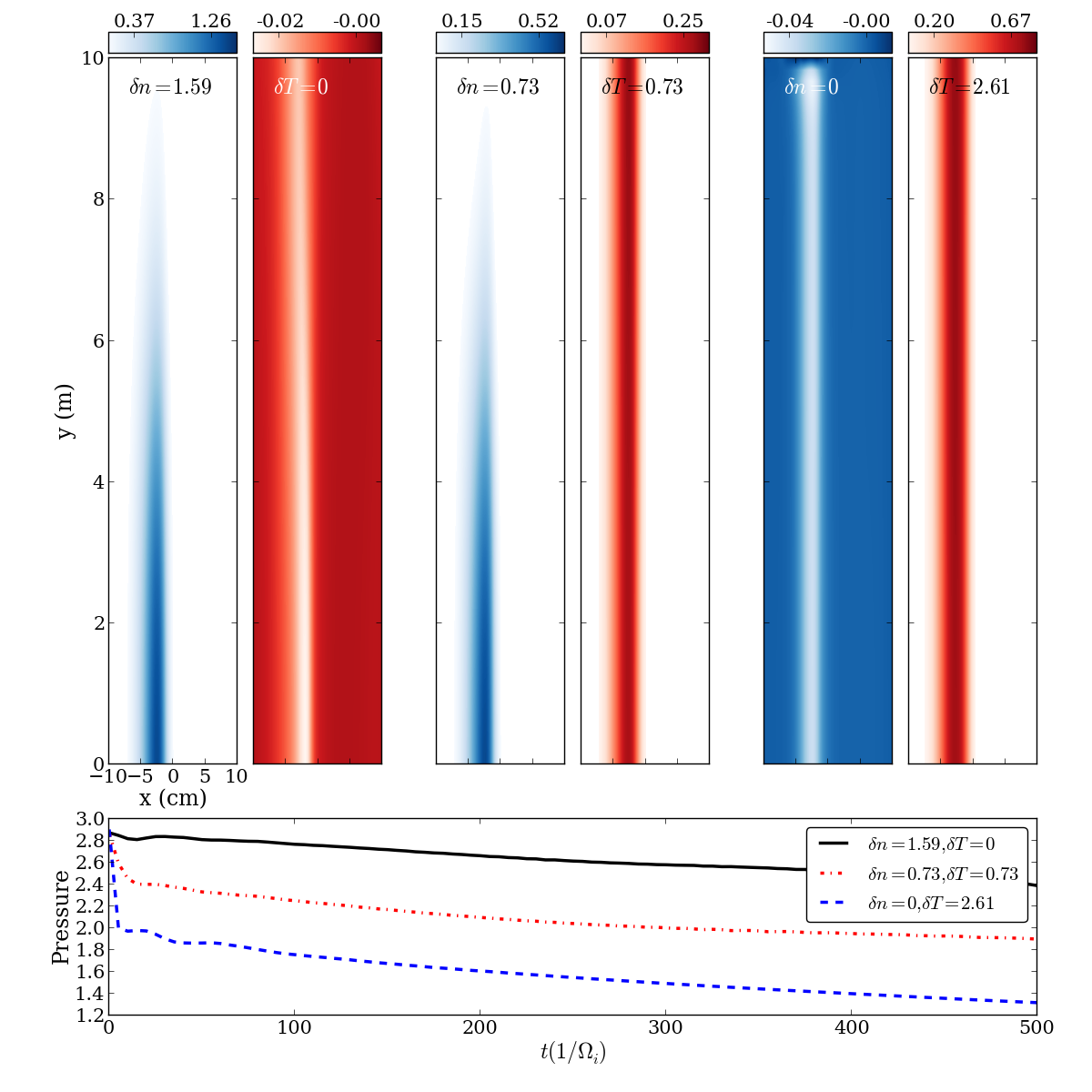}
\caption{Upper row: Parallel profile of the density and temperature (with background profiles subtracted off) across each set of amplitudes in the constant pressure scan. The profiles are taken at $t=150/\Omega_{i}$. Lower plot: Peak pressure in the filament a the midplane during the simulation. In the early phase a quick pressure drop occurs as electron conduction homogenises the temperature profile.}
\label{Fig:Par_Xsecs} 
\end{figure}

\begin{figure}[htbp]
\centering
\includegraphics[width=0.5\textwidth]{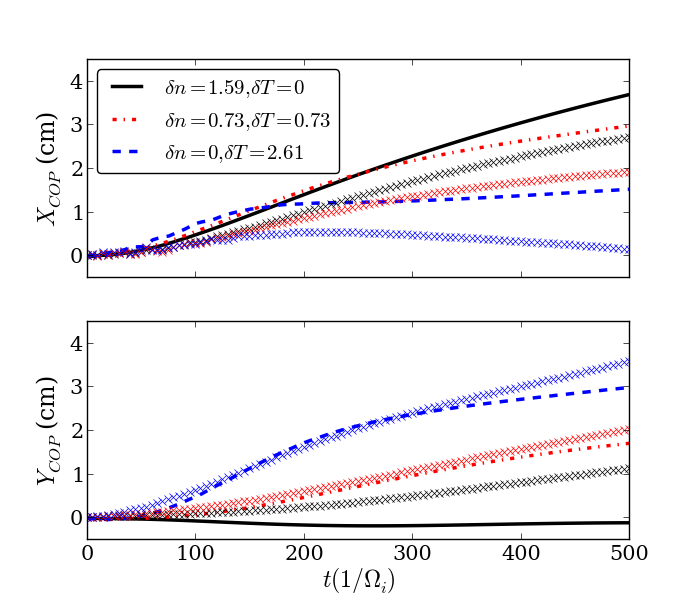}
\caption{COP coordinates in the radial (upper) and bi-normal (lower) direction during the filament evolution for filament amplitudes in the constant pressure scan. Crosses show the trajectory of filaments with a parallel lengthscale of half the domain (as shown in figure \ref{Fig:Par_Xsecs}) whilst lines show the trajectories of filaments with homogeneous parallel profiles.}
\label{Fig:COM_const_p_L5}
\end{figure}

Easy \emph{et al} \cite{EasyPoP2014} have recently shown that the mechanisms responsible for the motion of the filament can occur regardless of whether a physical connection to the sheath is present since an electrical connection can always be established through the plasma background. When significant parallel gradients are present in the filament however the Boltzmann response can result in net motion in the bi-normal direction \cite{AngusPoP2012,WalkdenPPCF2013,EasyPoP2014,WalkdenNF2015}. In thermal filaments the conductive heat flux is a much more efficient transport mechanism than parallel advection and as a result the parallel dynamics of a thermal filament are likely to vary with respect to the isothermal case. To test whether parallel profiles affect thermal dynamics simulations have been re-run with a parallel profile applied to the filament density and temperature of the form
\begin{equation}
\delta n,T = \delta n,T \left(x,y\right)\left(1 - \tanh\left(\frac{z - \delta_{z}}{0.3}\right)\right)
\end{equation}
with $\delta_{z} = 5$m. In figure \ref{Fig:Par_Xsecs} the parallel profile of the temperature and density is shown for each filament in the constant pressure scan at $t=150/\Omega_{i}$ after initialization. Also shown is the peak in the filament pressure at the midplane as a function of time.
\\Figure \ref{Fig:Par_Xsecs} shows that a thermal connection to the sheath is established in the filament by electron heat conduction, even though the density remains disconnected from the sheath. At the midplane this results in a quick loss of pressure on the thermal conduction timescale, followed by a more gradual loss once the filament has homogenised and sheath losses dominate. This allows the mechanisms leading to growth of $\phi^{+}$ to occur even when no physical connection to the target is present initially since the sheath is heated by the filament on a fast time scale. This is demonstrated in figure \ref{Fig:COM_const_p_L5} where the COP position in the radial and bi-normal direction are compared across the constant pressure scan in $\delta n$ and $\delta T$ between filaments with a parallel lengthscale of half the domain, and filaments with a homogeneous parallel profile.
For each combination of $\delta n$ and $\delta T$, the reduction in the filament parallel length scale reduces radial propagation, as observed in ref \cite{EasyPoP2014}. The net reduction in the radial propagation of the filament as the parallel lengthscale is varied is approximately constant indicating that the reduction is not affected by the manner in which the pressure is distributed between density and temperature. As noted already, the bi-normal motion is strongly influenced by the electron temperature and this is further evidenced in figure \ref{Fig:COM_const_p_L5}. The case with $\delta T = 0$ displays an increase in bi-normal propagation which occurs due to the Boltzmann response \cite{AngusPoP2012,WalkdenPPCF2013,EasyPoP2014,WalkdenNF2015} whilst the other two simulations show similar behaviour as the bi-normal motion is most strongly influenced by the temperature dependence of the floating potential.
\subsection{Effect of perpendicular lengthscale}
\begin{figure}[htbp]
\centering
\includegraphics[width=0.5\textwidth]{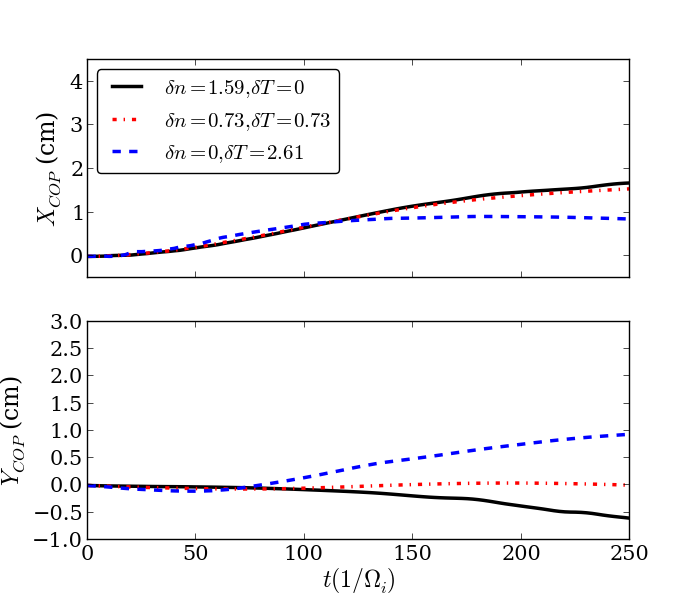}
\caption{COM coordinate in the radial (upper) and bi-normal (lower) direction for the filament in the inertial regime with $\delta_{\perp} = 0.5$cm across the constant pressure scan in $\delta n, \delta T$.}
\label{Fig:d05_COM_const_p}
\end{figure}

\begin{figure}[htbp]
\centering
\includegraphics[width=0.5\textwidth]{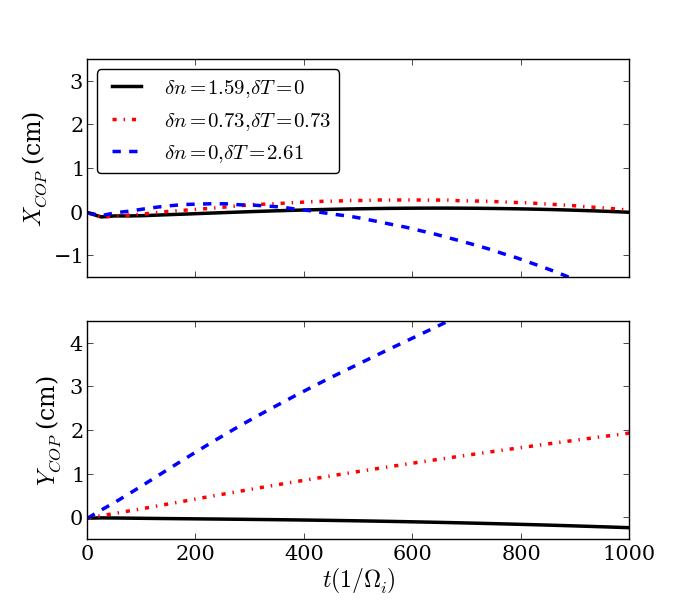}
\caption{COM coordinate in the radial (upper) and bi-normal (lower) direction for the sheath limited filament with $\delta_{\perp} = 8$cm across the constant pressure scan in $\delta n, \delta T$.}
\label{Fig:d8_COM_const_p}
\end{figure}

To test how the results presented so far depend on the radial size of the filament, two further simulation sets have been run with $\delta_{\perp} = 0.5cm < \delta^{*}$ and $\delta_{\perp} = 8cm > \delta^{*}$ respectively. These represent filaments in the inertial regime and in the sheath-limited regime. Figures \ref{Fig:d05_COM_const_p} and \ref{Fig:d8_COM_const_p} show the COP position for the simulation sets at each of these two radial sizes
\\The results of these simulations are broadly similar to those of the $\delta_{\perp} \sim \delta^{*}$ case. In both the $\delta_{\perp} = 0.5cm$ and $\delta_{\perp} = 8cm$ case there is a slowing of the filament compared with the $\delta_{\perp} = 2cm$ case, as predicted by the 2D velocity scaling. Within each constant pressure scan however the same trend is apparent: Increasing $\delta T$ at constant pressure leads to an increase in the bi-normal propagation and a decrease in the radial propagation of the filament. There are however some differences that are worth commenting on. First in the $\delta_{\perp} = 0.5cm$ case there is a net negative bi-normal motion when $\delta T = 0$. This is a dual effect of resistive drift waves that form towards the end of the simulation and start to break the filament down \cite{AngusPRL2012}, and the negative temperature perturbation that forms as the sheath heat flux increases locally driving a negative $\phi^{+}$. Second in the $\delta_{\perp} = 8cm$ case the radial motion reverses. For this sheath limited filament, the loss of pressure occurs faster than significant radial motion can occur. As such the filament ceases to exist properly, and the COP becomes dominated by negative perturbations that appears in the density field. The pressure loss is therefore more dominant in determining the motion of sheath limited filaments due to the small radial velocities and longer motion timescales than it is in the previous two cases. For inertially limited filaments and filaments with $\delta_{\perp} \sim \delta^{*}$ however the prominence of the even component of potential has the greater effect on determining the radial motion of the filament. Finally the increase in the bi-normal motion is much larger in the sheath limited case compared to the inertially limited case. This is purely a result of the larger size of the filament. Since $\phi^{e}$ develops a profile that overlaps the radial profile of the filament temperature it can move the filament further in the case then $\delta_{\perp} = 8$cm as compared to the case with $\delta_{\perp} = 0.5$cm. It is worth noting however that the net propagation as compared to $\delta_{\perp}$ remains similar in the two cases.

\section{Discussion}
Two main points of discussion arise from the results presented here. Firstly the analysis carried out is useful in establishing the region of validity of the commonly made isothermal assumption. The fundamental blob size has been shown to depend only very weakly on the ratio $(T_{0} + \delta T)/(n_{0} + \delta n)$ and the scalings of the inertial and sheath limited velocities on the perpendicular size of the blob deviate only weakly from the isothermal case. In both regimes however the velocities have differing dependencies on the temperature and density, meaning that they deviate from the isothermal case even if modifications are made to account for increased pressure in the filament. The isothermal assumption can therefore be considered adequate for establishing the transition point from the inertial to sheath limited regime in terms of $\delta_{\perp}$, but cannot capture in detail the magnitude of the velocity properly.  
\\The same cannot be said when making a detailed investigation of the filament dynamics. The presence of the sheath potential in the thermal case has been shown to induce filament motion in the bi-normal direction. This occurs even when no physical connection to the sheath is present due to efficient thermal conduction quickly establishing a thermal connection to the sheath. The radial motion is also affected by the filament temperature. In the early stages the motion is relatively constant at constant pressure, in line with velocity scaling predictions, however the long term propagation is reduced by the circulatory motion and asymmetry in the filament cross-section that occurs when $\delta T$ is large. This implies that filaments with $\delta T/T_{0} >> \delta n/ n_{0}$ may be limited to the region close to the separatrix whilst filaments with $\delta T/T_{0} << \delta n/ n_{0}$ may propagate further into the SOL. The isothermal assumption can therefore be reasonably applied when modelling filaments that propagate into the far SOL, but could be inaccurate when applied to filaments that stay in the near SOL. 
\\The second point of discussion concerns the role played by filaments in power loading to the divertor. The results presented here have shown that the radial propagation of a filament is dependant on the relative weighting of their density and temperature perturbations, with larger $\delta T/T_{0}$ relative to $\delta n/n_{0}$ leading to reduced radial propagation. This implies that heat carried into the SOL by filaments is more confined near to the separatrix since filaments with large $\delta T/T_{0}$ (and therefore carrying significant heat) require an even larger $\delta n/n_{0}$, which will be a rare event, to propagate significantly far from the separatrix. This implies that the deposition of heat onto the divertor will be more peaked in the near-SOL region than expected when radial transport is considered constant across the SOL and peaking is purely due to parallel transport processes. It has recently been shown on MAST that filaments can play a crucial role in determining the heat flux to divertor components \cite{ThorntonPPCF2015,ThorntonPrePrint2016} so understanding how their motion and heat deposition are related may be an important step in predicting divertor footprints due to filaments.

\section{Conclusions}
This paper has presented a numerical investigation of the role  of thermal effects in 3D isolated filament dynamics. A series of simulations have been run with a fixed pressure within the filament thereby fixing the drive for the filament motion, but with varying density and temperature amplitudes, thereby isolating the role of thermal effects. Simulations have been carried out for filaments in the inertial and sheath limited regimes and for filaments between the two. For each different filament size a similar trend is apparent with the following two factors consistent between all simulations; filaments with $\delta T/T_{0} >> \delta n/n_{0}$ show an increase in their propagation in the bi-normal direction and a reduction in propagation in the radial direction. The bi-normal motion is driven by the sheath current, which provides a drive for an even parity component of the electrostatic potential (about the filament centre in the bi-normal direction) which, when coupled with the odd parity component arising from the curvature drive, leads to bi-normal motion. This occurs even when there is no thermal connection to the sheath in the initial state due to the fast conduction of the electron temperature along the field line. The rotational motion driven by the even component of the potential recirculates the filament which acts against the radial motion and reduces the net radial heat and particle flux from the filament. Importantly this can only occur when the angular symmetry of the filament cross-section is broken. This reduction in radial propagation is supplemented by the comparatively fast loss of pressure in the filament when $\delta T/T_{0} >> \delta n/n_{0}$. By quickly reducing the pressure the drive is reduced for the radial propagation of the filament. This is observed to be the dominant process limiting the radial propagation of sheath limited filaments.
\\The observations made here suggest that filaments with weak temperature perturbations are able to propagate further into the SOL than those with large temperature perturbations (at constant pressure) and may result in more peaked heat flux to the divertor than just the peaking provided by parallel transport. 

\section{Acknowledgements}
This work was part-funded by the RCUK Energy Programme under grant EP/I501045 and the European Communities. Simulations in this paper made use of the ARCHER UK National Supercomputing service (www.archer.ac.uk) under the Plasma HEC Consortium EPSRC grant number EP/L000237/1. To obtain further information on the data and models underlying this paper please contact PublicationsManager@ccfe.ac.uk. The views and opinions expressed herein do not necessarily reflect those of the European Commission.

\section{References}
\bibliographystyle{prsty}
\bibliography{Bibliography}

\section{Appendix A: 1D Background}
\label{App:1D_eqm}
\begin{figure}[htbp]
\centering
\includegraphics[width=0.7\textwidth]{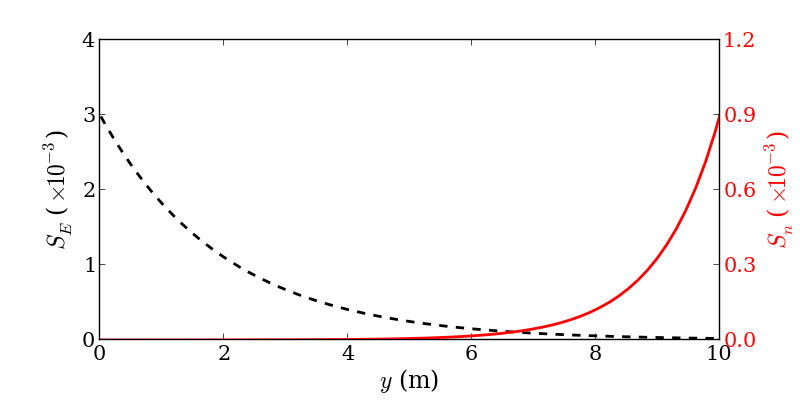}
\caption{Source profiles used to create backgrounds for simulations run in this paper. The energy source (black broken line, left axis) and the particle source (red solid line, right axis) are shown as a function of parallel distance with $y=0$ at the midplane and $y=L_{||}=10$m at the sheath.}
\label{Fig:Sources}
\end{figure}

\begin{figure}[htbp]
\centering
\includegraphics[width=0.7\textwidth]{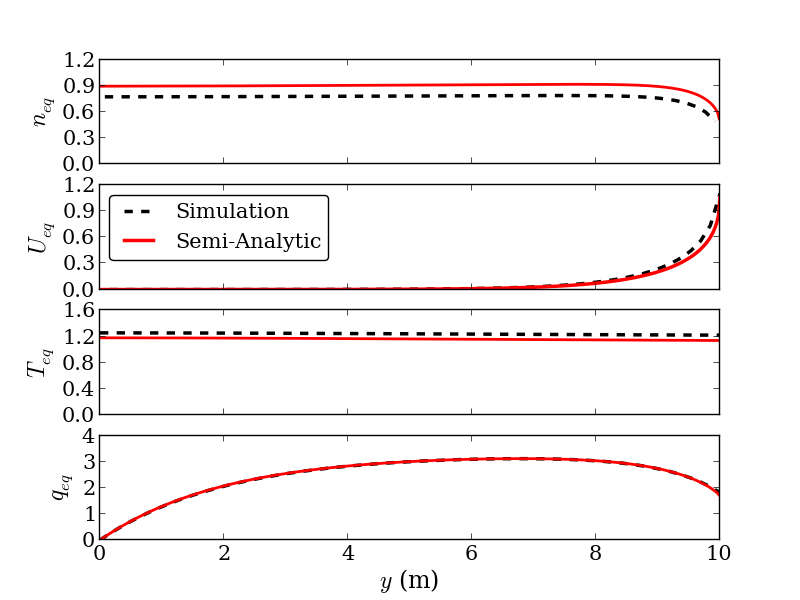}
\caption{Background equilibrium profiles calculating directly from simulation (black broken line) and semi-analytically (red solid line).}
\label{Fig:eq_comp}
\end{figure}

Following the approach of Easy \emph{et al} \cite{EasyPoP2014} filaments are seeded on top of pre-determined background equilibrium profiles with variation in the field aligned-direction only. These profiles are determined by the equilibrium solution of equations (\ref{Eqn:Dens}) to (\ref{Eqn:Temp}) neglecting all perpendicular variation. The equilibrium profiles are determined by the particle and energy source profiles, $S_{n}$ and $S_{E}$, which balance with sinks at the target boundary. These profiles are shown in figure \ref{Fig:Sources}.
\\The source profiles are chosen to represent a density source dominated by recycling close to the divertor plate and an energy source arising from power crossing the separatrix ballooned around the outboard midplane. Whilst the recycling source of density may be approximately volumetric close to the divertor plates, it is unlikely that the source of energy at the midplane will be volumetric and may not be well represented by a 1D profile. The introduction of a radially varying background profile, whilst resembling experiment more closely, will add extra complexity to the simulations. Since the aim of this paper is to understand the role of $T_{e}$ on filament dynamics on a fundamental level this extra complexity is reserved for future study. Instead  a basic understanding of the role played by $T_{e}$ on the motion of filaments is sought, for which 1D background profiles are sufficient. 
\\The 1D background profiles are produced by running the same module used for filament simulations to saturation with no perpendicular variation included. The resulting profiles  have been compared to a semi-analytically derived set of profiles. These semi-analytic profiles are obtained for $n$ and $V$ using the analytic expressions derived in \cite{EasyPoP2014} with the addition of temperature variation giving
\begin{equation}
\label{Eqn:n_eq}
n_{eq} = \frac{1}{V_{eq}}\int_{0}^{y}S_{n}\left(y'\right)dy'
\end{equation}
and 
\begin{equation}
\label{Eqn:n_eq}
V_{eq} = \frac{1+\beta - \sqrt{\left(1+\beta\right)^{2} - 4T_{eq}\alpha^{2}\beta}}{2\alpha\beta}
\end{equation}
with $\beta$ and $\alpha$ defined by
\begin{equation}
\alpha = \frac{\int_{0}^{y}S_{n}\left(y'\right)dy'}{\sqrt{T_{sh}}\int_{0}^{L_{||}}S_{n}\left(y'\right)dy'}
\end{equation}
where $T_{sh}$ is the temperature at the sheath and
\begin{equation}
\beta = 1 + \frac{m_{e}}{m_{i}}
\end{equation}
These expressions depend directly or indirectly on the equilibrium temperature profile, for which no analytic expression has been found. Instead Newton-Krylov iteration (using the krylov solver in the \texttt{root} function from the \texttt{optimize} module of the python \texttt{scipy} package) has been used to solve the 1D equilibrium temperature equation,
\begin{equation}
\label{Eqn:Eqm_T}
\frac{3}{2}n_{eq}V_{eq}\nabla_{||}T_{eq} + \nabla_{||}q_{||,eq} + n_{eq}T_{eq}\nabla_{||}V_{eq} = S_{E} + \frac{1}{2\mu}V_{eq}^{2}S_{n} - \frac{3}{2}T_{eq}S_{n}
\end{equation}
Spatial derivatives are calculated using a 4th order central difference method. Figure \ref{Fig:eq_comp} compares BOUT++ equilibrium profiles with the semi-analytically derived profiles using the sources shown in figure \ref{Fig:Sources}.
Some minor deviation between the BOUT++ and semi-analytic profiles is observable in figure \ref{Fig:eq_comp} which can be attributed to numerical dissipation in the BOUT++ solution. This causes the density profile to reduce compared to the analytic profile, as observed in ref. \cite{EasyPoP2014}, which in turn impacts the $q_{||}$ boundary condition and causes slight variation in other quantities, however such minor variation does not impact the simulations.
\\The profiles shown in figure \ref{Fig:eq_comp} have been designed to be comparable to the isothermal profiles described in ref \cite{EasyPoP2014} and are in the sheath limited regime of divertor operation \cite{Stangeby}. 

\end{document}